\documentclass[preprintnumbers,article,amsmath,amssymb,floatfix,10pt,prd,onecolumn,
superscriptaddress,nofootinbib]{revtex4-2}
\usepackage{bm}
\usepackage{amsfonts}
\usepackage{latexsym}
\usepackage{graphicx}
\usepackage{amsmath}
\usepackage{textcomp}
\usepackage{hyperref}
\hypersetup{colorlinks,citecolor=blue}
\hypersetup{colorlinks=true,linkcolor=red,filecolor=magenta,    urlcolor=blue}
\usepackage{amsmath}
\usepackage{xcolor}
\usepackage{epsfig}

\newcommand{\m}{\mathring}
\newcommand{\eea}{\end{eqnarray}}
\newcommand{\eeas}{\end{eqnarray*}}

\def\jnl@style{\it}
\def\aaref@jnl#1{{\jnl@style#1}}

\def\aaref@jnl#1{{\jnl@style#1}}

\def\aj{\aaref@jnl{AJ}}                   
\def\apj{\aaref@jnl{ApJ}}                 
\def\apjl{\aaref@jnl{ApJ}}                
\def\apjs{\aaref@jnl{ApJS}}               
\def\apss{\aaref@jnl{Ap\&SS}}             
\def\aap{\aaref@jnl{A\&A}}                
\def\aapr{\aaref@jnl{A\&A~Rev.}}          
\def\aaps{\aaref@jnl{A\&AS}}              
\def\mnras{\aaref@jnl{Mon.~Not.~Roy.~Astron.~Soc.}}             
\def\prd{\aaref@jnl{Phys.~Rev.~D}}        
\def\prc{\aaref@jnl{Phys.~Rev.~C}}  
\def\prl{\aaref@jnl{Phys.~Rev.~Lett.}}    
\def\qjras{\aaref@jnl{QJRAS}}             
\def\skytel{\aaref@jnl{S\&T}}             
\def\ssr{\aaref@jnl{Space~Sci.~Rev.}}     
\def\zap{\aaref@jnl{ZAp}}                 
\def\nat{\aaref@jnl{Nature}}              
\def\aplett{\aaref@jnl{Astrophys.~Lett.}} 
\def\apspr{\aaref@jnl{Astrophys.~Space~Phys.~Res.}} 
\def\physrep{\aaref@jnl{Phys.~Rep.}}      
\def\physscr{\aaref@jnl{Phys.~Scr}}       
\def\commat{\aaref@jnl{Comm.~Math.~Phys.}}              
\def\science{\aaref@jnl{Science}}               
\def\cqg{\aaref@jnl{Classical Quant.~Grav.}}            
\def\jpcs{\aaref@jnl{JPCS}}                                     
\def\ijmpd{\aaref@jnl{Int.~J.~Mod.~Phys.~D}}                    
\def\grg{\aaref@jnl{Gen.~Relat.~Gravit.}}               
\def\rpp{\aaref@jnl{Rep.~Prog.~Phys.}}          
\def\npa{\aaref@jnl{Nucl.~Phys.~A}}        
\def\lrr{\aaref@jnl{Living Rev.~Rel.}}                   
\def\jcap{\aaref@jnl{J.~Cosmology Astropart.~Phys.}}    
\def\rmp{\aaref@jnl{Rev.~Mod.~Phys.}}   
\def\epjc{\aaref@jnl{Eur.~Phys.~J.~C}} 
\def\plb{\aaref@jnl{~Phy.~Lett.~B}} 
\def\mpla{\aaref@jnl{Mod.~Phy.~Lett.~A}} 
\def\arxiv{\aaref@jnl{arxiv.org}}

\allowdisplaybreaks[1]

\begin{document}

\title{ Cosmology of $f(Q)$ gravity in non-flat Universe }

\author{Hamid Shabani}
\affiliation{Physics Department, Faculty of Sciences, University of Sistan and 
Baluchestan, Zahedan, Iran}

\author{Avik De}
\address{Department of Mathematical and Actuarial Sciences\\
Universiti Tunku Abdul Rahman, Jalan Sungai Long, 43000 Cheras, Malaysia}

\author{Tee-How Loo}
\affiliation{Institute of Mathematical Sciences, Faculty of Science, Universiti 
Malaya, 50603 Kuala Lumpur, Malaysia}

\author{Emmanuel N. Saridakis}
\affiliation{National Observatory of Athens, Lofos Nymfon, 11852 Athens, 
Greece}
\affiliation{CAS Key Laboratory for Researches in Galaxies and Cosmology, 
Department of Astronomy, University of Science and Technology of China, Hefei, 
Anhui 230026, P.R. China.}
\affiliation{Departamento de Matem\'{a}ticas, Universidad Cat\'{o}lica del 
Norte,  Avda. Angamos 0610, Casilla 1280 Antofagasta, Chile}

\begin{abstract}
We investigate the cosmological implications of $f(Q)$ gravity, which is a 
modified theory of gravity based on non-metricity, in non-flat   geometry. We 
perform a detailed dynamical-system  analysis keeping the $f(Q)$ function  
completely arbitrary. As we show, the cosmological scenario admits a dark-matter 
dominated point, as well as a dark-energy dominated  de Sitter solution which   
can attract the Universe at late times. However, the main result of the present 
work is that there are additional critical points  which exist solely due to 
curvature. In particular, we find that there are curvature-dominated 
accelerating points  which are unstable and thus   can describe the inflationary 
epoch. Additionally, there is a point in which the dark-matter and dark-energy 
density parameters are both between zero and one, and thus  it  can alleviate 
the coincidence problem. Finally, there is a saddle point which is completely 
dominated by curvature. In order to provide a specific example, we apply our 
general analysis to the power-law case, showing that we can obtain the thermal 
history of the Universe, in which  the curvature density parameter may exhibit 
a peak at intermediate times.   These features, alongside 
possible indications that non-zero curvature could alleviate the cosmological 
tensions, may serve as advantages for $f(Q)$ gravity in non-flat geometry.

\end{abstract}
\maketitle

\section{Introduction}\label{sec1}

Modified gravity \cite{CANTATA:2021ktz,Capozziello:2011et}
is one of the 
two main directions that one can follow in order to obtain an improved 
description of the Universe evolution, both concerning the early (inflation) and 
late (dark-energy) accelerated phases, as well as concerning the possible 
observational  tensions \cite{Abdalla:2022yfr}. In such theories one constructs 
modifications and extensions of General Relativity which present extra degrees 
of freedom capable of inducing corrections at the cosmological behavior, both 
at the background and perturbation level.

There are many ways to construct gravitational modifications. In the simplest 
ones one starts from the Einstein-Hilbert Lagrangian and adds new terms, 
resulting to    $f(R)$ gravity 
\cite{Starobinsky:1980te},  to  $f(G)$
gravity \cite{Nojiri:2005jg}, to  $f(P)$ 
gravity 
\cite{Erices:2019mkd}, to Lovelock gravity \cite{Lovelock:1971yv},   to
Horndeski/Galileon scalar-tensor theories 
\cite{Horndeski:1974wa,Deffayet:2009wt}  etc. Alternatively, one may start from 
the torsion-based formulation of gravity and modify it accordingly, resulting 
to   $f(T)$ gravity 
\cite{Bengochea:2008gz,Cai:2015emx},   $f(T,T_{G})$ gravity 
\cite{Kofinas:2014owa},   $f(T,B)$ gravity 
\cite{Bahamonde:2015zma},
  scalar-torsion theories \cite{Geng:2011aj} etc.

One different class of gravitational modifications arises when one starts 
from the equivalent formulation of gravity based on non-metricity. Initiated by 
Nester and Yo \cite{Nester:1998mp}, based on an affine 
connection with vanishing curvature and torsion but metric-incompatibility, it 
was recently extended to $f(Q)$ theory 
\cite{coincident}.  $f(Q)$     gravity  contains general relativity  as a 
particular 
limit, and  has the advantage of possessing 
second-order field equations. Hence, its cosmological application has attracted 
the interest of the  literature
\cite{cosmology,cosmography,barros,lu,Anagnostopoulos:2021ydo,de/comment,de/iso,
Solanki:2022rwu,
Solanki:2022ccf,
Beh:2021wva,cosmology_Q,signature,lcdm1,siren,recon,recon1,anisotropy,
De:2022vfc,lazkos2019,Ayuso:2020dcu,bajardi2020,
Lymperis:2022oyo,sahoo2022,DAmbrosio:2021zpm,Li:2021mdp,Dimakis:2021gby,
Hohmann:2021ast, 
Kar:2021juu,Wang:2021zaz,Quiros:2021eju,Mandal:2021bpd,Albuquerque:2022eac, 
Papagiannopoulos:2022ohv,Anagnostopoulos:2022gej,Arora:2022mlo,Pati:2022dwl, 
Maurya:2022wwa,Capozziello:2022tvv,Dimakis:2022wkj,DAgostino:2022tdk,
Narawade:2022cgb,Emtsova:2022uij,Bahamonde:2022cmz,Narawade:2023nzv,
Ferreira:2023tat,Sokoliuk:2023ccw,Shaikh:2023tii,Jan:2023djj,Dimakis:2023uib,
Koussour:2023rly,Najera:2023wcw,Atayde:2023aoj}. Nevertheless, all of these 
works   focus on   spatially-flat 
Friedmann-Lema\^itre-Robertson-Walker (FLRW) geometry, in which case   the 
coincident gauge   implies that  the affine connection field equations can 
be ignored   and thus $f(Q)$ cosmology coincides with $f(T)$ cosmology at the 
background level \cite{Jarv:2018bgs}.  

In this work we are interested in investigating $f(Q)$ cosmology in non-flat 
Universe, in order to reveal possible novel features, having in mind that 
non-flat geometry \cite{ellis}, apart from being potentially interesting 
\cite{yang2022,pan,Subramaniam:2023old,holo,valentino2020,vagnozzi2021,
Vagnozzi:2020dfn, dhawan2021,perturbclosed,glanville2022,fTpert},  might 
be one way to alleviate cosmological tensions \cite{DiValentino:2019qzk}. The 
manuscript is organized as follows: In Section~\ref{sec2} we provide the basic 
mathematical formalism of symmetric teleparallel and $f(Q)$ gravity. Then, in 
Section \ref{sec4} we  perform a detailed dynamical-system analysis, extracting 
the general cosmological features, both for a general $f(Q)$ function as well 
as for a specific power-law example. Finally, in Section~\ref{sec8} we 
summarize our results.


\section{Symmetric teleparallel and $f(Q)$ gravity}
\label{sec2}

In this section we briefly review symmetric teleparallel formulation of gravity 
and its $f(Q)$ extension.  In such a formalism one introduces a 
general affine connection $\Gamma^\alpha_{\,\,\, \beta\gamma}$, defined by 
$
\Gamma^\lambda{}_{\mu\nu} = 
\mathring{\Gamma}^\lambda{}_{\mu\nu}+L^\lambda{}_{\mu\nu},
$
where $\mathring{\Gamma}^\lambda{}_{\mu\nu}$ is the Levi-Civita connection and 
the disformation tensor  is given by
\begin{equation} \label{L}
L^\lambda{}_{\mu\nu} = \frac{1}{2} (Q^\lambda{}_{\mu\nu} - 
Q_\mu{}^\lambda{}_\nu - Q_\nu{}^\lambda{}_\mu) \,,
\end{equation}
with   the non-metricity tensor given as
\begin{equation} \label{Q tensor}
Q_{\lambda\mu\nu} = \nabla_\lambda g_{\mu\nu} \,.
\end{equation}
Additionally, one can define the non-metricity scalar as
\begin{equation} \label{Q}
Q=Q_{\lambda\mu\nu}P^{\lambda\mu\nu}= 
\frac{1}{4}(-Q_{\lambda\mu\nu}Q^{\lambda\mu\nu} + 
2Q_{\lambda\mu\nu}Q^{\mu\lambda\nu} +Q_\lambda Q^\lambda -2Q_\lambda 
\tilde{Q}^\lambda),
\end{equation}
with $
Q_{\lambda}=Q_{\lambda\mu\nu}g^{\mu\nu}$ and $  \tilde 
Q_{\nu}=Q_{\lambda\mu\nu}g^{\lambda\mu}$. Hence, using $Q$ as a Lagrangian 
gives rise to the same equations with general relativity.

Based on the symmetric teleparallel framework, one can proceed in 
constructing gravitational modifications, such as $f(Q)$ gravity 
\cite{coincident}, characterized by the action  
\begin{equation}
S = \frac1{2\kappa}\int f(Q) \sqrt{-g}\,d^4 x
+\int \mathcal{L}_M \sqrt{-g}\,d^4 x,
\end{equation}
where we have added the matter Lagrangian for completeness, corresponding to 
an energy-momentum tensor  of a perfect fluid 
$
T^{m}_{\mu\nu}=(p+\rho)u_\mu u_\nu+pg_{\mu\nu}$,
with $p$ and $\rho$   the pressure and energy density respectively. Variation 
of the action with respect to the metric leads to the field equations
\begin{equation} \label{FE1}
\frac{2}{\sqrt{-g}} \nabla_\lambda (\sqrt{-g}FP^\lambda{}_{\mu\nu})  
-\frac{1}{2}f g_{\mu\nu} + F(P_{\nu\rho\sigma} Q_\mu{}^{\rho\sigma} 
-2P_{\rho\sigma\mu}Q^{\rho\sigma}{}_\nu) = \kappa T^{m}_{\mu\nu},
\end{equation}
where the superpotential $P^\lambda{}_{\mu\nu}$ is given by
\begin{equation} \label{P}
P^\lambda{}_{\mu\nu} = \frac{1}{4} \left( -2 L^\lambda{}_{\mu\nu} + Q^\lambda 
g_{\mu\nu} - \tilde{Q}^\lambda g_{\mu\nu} -\frac{1}{2} \delta^\lambda_\mu 
Q_{\nu} - \frac{1}{2} \delta^\lambda_\nu Q_{\mu} \right) \,,
\end{equation}
and with $F(Q)=df(Q)/dQ$. 
Note that the field equations (\ref{FE1}) can be alternatively written as 
  \cite{zhao}
\begin{equation} \label{FE}
F \m{G}_{\mu\nu}+\frac{1}{2} g_{\mu\nu} (FQ-f) + 2F' P^\lambda{}_{\mu\nu} 
\m{\nabla}_\lambda Q = \kappa T^{m}_{\mu\nu},
\end{equation}
where $\mathring{G}_{\mu\nu} = \mathring{R}_{\mu\nu} - \frac{1}{2} g_{\mu\nu} 
\mathring{R}$, and all the expressions denoted with a $\mathring{()}$ are 
calculated with respect to the Levi-Civita connection 
$\mathring{\Gamma}^\lambda{}_{\mu\nu}$. Hence, we can re-write them as   
\cite{de/comment}
\begin{equation}\label{equiv}
    \mathring{G}_{\mu\nu}= \frac{\kappa}{F}T^{m}_{\mu\nu}+\kappa 
T^{\text{de}}_{\mu\nu},
\end{equation}
having defined an effective dark-energy sector of geometrical origin as
\begin{equation}
\kappa 
T^{\text{de}}_{\mu\nu}=\frac{1}{F}\left[\frac{1}{2}g_{\mu\nu}(f-QF)-2F'\mathring
{\nabla}_\lambda QP^\lambda_{\mu\nu}\right],
\label{Tde}
\end{equation}
where a prime denotes differentiations with respect to the argument. Lastly, 
varying the action   with respect to the affine connection, and assuming 
that   the matter Lagrangian $\mathcal{L}_{M}$ does not depend on it, we obtain
\begin{align}\label{FE2}
\nabla_\mu\nabla_\nu(\sqrt{-g}F P^{\nu\mu}{}_\lambda)=0\,.
\end{align}


Let us apply $f(Q)$ gravity to a cosmological framework. As we mentioned in the 
Introduction, we   consider a non-flat  Friedmann-Lema\^itre-Robertson-Walker 
(FLRW)  spacetime of the form 
\begin{align}\label{metric}
ds^2 = - d t^2 
+a\left(t\right)^{2}\left( \frac{dr^2}{1-kr^2} 
+r^2\mathrm{d}\theta^2+r^2\sin^2\theta\mathrm{d} \phi^2\right),  
\end{align}
where $k=0,\,\pm1$ denotes the spatial curvature.  In this case, the 
non-trivial connection coefficients are given by \cite{FLRW/connection}
\begin{align}\label{aff}
\Gamma^t{}_{tt}=&-\frac{k+\dot\gamma}\gamma, 
	\quad 					\Gamma^t{}_{rr}=\frac{\gamma}{1-kr^2}, 
	\quad 					\Gamma^t{}_{\theta\theta}=\gamma r^2, 
	\quad						\Gamma^t{}_{\phi\phi}=\gamma r^2\sin^2\theta	
			 				\notag\\
\Gamma^r{}_{tr}=&-\frac{k}{\gamma}, 
	\quad  	\Gamma^r{}_{rr}=\frac{kr}{1-kr^2}, 
	\quad		\Gamma^r{}_{\theta\theta}=-(1-kr^2)r, 
	\quad		\Gamma^r{}_{\phi\phi}=-(1-kr^2)r\sin^2\theta,					
							 \notag\\
\Gamma^\theta{}_{t\theta}=&-\frac{k}{\gamma}, 
	\quad		\Gamma^\theta{}_{r\theta}=\frac1r,
	\quad		\Gamma^\theta{}_{\phi\phi}=-\cos\theta\sin\theta,				
					 	\notag\\
\Gamma^\phi{}_{t\phi}=&-\frac k\gamma,  
	\quad 	\Gamma^\phi{}_{r\phi}=\frac1r, 
	\quad 	\Gamma^\phi{}_{\theta\phi}=\cot\theta,
\end{align}
where $\gamma(t)$ is  a non-zero function of time. 
The corresponding non-metricity scalar $Q$  can be calculated from (\ref{aff})  
as \cite{FLRW/connection,FLRW/connection1,De:2022jvo,Paliathanasis:2023nkb} 
\begin{equation}\label{Q}    
Q(t)=-3\left[2H^2+\left(\frac{3k}{\gamma}-\frac{\gamma}{a^2}\right)H-\frac{2k}{
a^2}-k\frac{\dot{\gamma}}{\gamma^2}-\frac{\dot{\gamma}}{a^2}\right].
\end{equation}
Therefore, inserting into the 
 field equations (\ref{FE}) we obtain the modified Friedmann   equations
\begin{align}\label{rho} 
 -\frac12f-\left(3H^2+3\frac k{a^2}-\frac12Q\right)F+\frac32\dot 
Q\left(\frac k\gamma+\frac\gamma{a^2}\right)F'+\kappa \rho^{m}=0 
\end{align}
\begin{align}\label{p}
\frac12f+\left(3H^2+2\dot H+\frac k{a^2}-\frac12Q\right)F
        +\dot Q\left(2H+\frac32\frac 
k\gamma-\frac12\frac\gamma{a^2}\right)F'+\kappa p^{m}=0.
\end{align}

%
 
\section{Cosmological behavior}
\label{sec4}

In this section we investigate in detail the cosmological evolution of a 
Universe governed by $f(Q)$ gravity in   non-flat geometry. In order to achieve 
that we perform a dynamical-system analysis  
\cite{Coley:2003mj,Bahamonde:2017ize}, which  allows one to 
extract the global features of a cosmological scenario independently of the 
specific initial conditions 
\cite{Copeland:1997et,	
 Setare:2008sf, Matos:2009hf, 
Copeland:2009be, Leon:2013qh,
 Skugoreva:2014ena, Hernandez-Almada:2021rjs}. Note that the dynamical-system 
analysis for $f(Q)$ gravity has been performed in the literature 
\cite{Narawade:2022jeg,Khyllep:2022spx,Boehmer:2023knj, 
Narawade:2023vpf,Shabani:2023nvm}, however it remains in the flat FLRW case, 
while as we will see in the following the inclusion of spatial curvature leads 
to novel qualitative features. 
 
 We start   by defining   dimensionless variables  in order  to
re-write Eqs.~(\ref{rho})-(\ref{p}) as an autonomous system. For simplicity we 
will focus on dust matter, namely we consider 
 $p^{m}=0$, while concerning the $\gamma(t)$ form we assume the 
simple case $\gamma(t)=\epsilon a(t)$ (analysis  of the general 
case is straightforward, with the inclusion of  an extra variable 
\cite{Shabani:2023nvm}). In particular, we have
\begin{align} 
&x_1=-\frac{f}{6H^{2}F},~~~~~x_2=\frac{Q}{6H^{2}},~~~~~x_3=\frac{\dot{F}}{HF},
~~~~~x_4=\frac{1}{2Ha},\nonumber\\
&\Omega^{m} =\frac{\kappa\rho^{m}}{3H^{2}F},~~~~~~\Omega^{k} 
=-\frac{k}{H^{2}a^{2}},\nonumber\\
&r=-\frac{QF}{f}=\frac{x_2}{x_1},~~~m=\frac{QF'}{F},\label{ds1bb}
\end{align}
where $\Omega^{m}$ and $\Omega^{k}$ denote the contributions to the dark matter 
and the spatial curvature energy densities, and thus according to (\ref{Tde}) 
  the first Friedmann equation (\ref{rho}) is written as 
$1=\Omega^{m}+\Omega^{k}+\Omega^{de}$ . The two parameters $m$ and $r$ 
parametrize the $f(Q)$ form as a function $m(r)$, while the variable $x_4$ 
has been introduced in order to break the degeneracy between positive and 
negative $H$ (since it is $H^2$ that appears in the other variables).

Using the above dimensionless variables we result to the   four-dimensional 
 autonomous system  
\begin{align}
&\frac{dx_1}{dN}=-\frac{x_2x_3}{m}-x_1(x_3+3\mathcal{A}),\label{ds7}\\
&\frac{dx_2}{dN}=\frac{x_2x_3}{m}-3x_2\mathcal{A},\label{ds8}\\
&\frac{dx_3}{dN}=-x_3\left(3+x_3+\frac{3}{2}\mathcal{A}\right),\label{ds9}\\
&\frac{\Omega^{k}}{dN}=-\Omega^{k}(2+3\mathcal{A}),\label{ds10}
\end{align}
with $\mathcal{A}  
\equiv \frac{2\dot{H}}{3H^{2}}=-1+x_1+x_2+\frac{x_3}{3}\left(\zeta 
x_4-2\right)+\frac{\Omega^k}{3}$ and $\zeta=-3\frac{k}{\epsilon}+\epsilon$,
and where   $N=\ln a$. Hence, the total equation-of-state parameter is just 
 $w^{eff}=-1-2\dot{H}/{3H^2}=-1-\mathcal{A}$. 
 Finally, we mention here that since $r=x_2/x_1$, the condition 
$dr/dN=rx_3\left(1+\frac{1+r}{m}\right)=0$   implies that the critical points 
  of the system   (\ref{ds7})-(\ref{ds10}) must  
satisfy either   $r=0$ (or equivalently $x_2=0$), or $x_3=0$, or $m(r)=-(1+r)$, 
while  when the conditions $x_1=0$, $x_2=0$ and $x_3\neq0$ 
simultaneously hold  the relation $m(r)=-(1+r)$ must be considered.

\subsection{General $f(Q)$ form}\label{sec5}

We start by performing the analysis for a general $f(Q)$ form, namely for a 
general   $m(r)$ function. 
As we will see, in this case 
 the intersections of the curve $m(r)$ with the line $m=-r-1$ can 
play an important role in the way that the critical point  corresponding to   
dark-matter dominated era connects to those exhibiting   dark-energy 
domination.  

In the general  case the  critical points of the system 
(\ref{ds7})-(\ref{ds10}) are presented in Table~\ref{tab1}. As can be seen, for 
a general $f(Q)$ function there exist seven critical points, or curves of 
critical points, with different physical features. Note that  
all $m$ and $m'$ values must be calculated at probable intersections of $m(r)$ 
with $m=-r-1$, which happen at the roots $r_i,~i=1,2,\cdots$.

\begin{center}
\begin{table}[h]
\centering
\begin{tabular}{l @{\hskip 0.1in} l@{\hskip 0.1in} l @{\hskip 0.1in}l @{\hskip 
0.1in}l @{\hskip 0.1in}l}\hline\hline
Fixed point     &Coordinates $(x_1,x_2,x_3,\Omega^{k})$           &Eigenvalues  
&$\Omega^m$   &$\Omega^k$   &$w^{eff}$\\[0.5 ex]
\hline
$P^{m}$&$\left(0,0,0,0)\right)$&$\left[3, 3, -\frac32, 
1\right]$&$1$&$0$&$0$\\[0.75 ex]
$P^{k}$&$\left(0,0,0,1\right)$&$\left[-2,-1,2,2\right]$&$0$&$1$&$-\frac13$\\[
0.75 ex]
$P^{ds}$&$\left(x_1,1-x_1,0,0\right)$   &$[-3, -2, 0, 
-3]$&$0$&$0$&$-1$\\[0.75 ex]
 $P^{1}$&$\left(0,0,-2,-\frac{\epsilon 
^2}{k}\right)$ with $k\neq0$ &$\left\{\begin{array}{l}\left[1,2,2-\frac{2}{m},2 
\left(\frac{1}{m}+1\right) m'+4\right],\\k=-\epsilon ^2,\\\left\{2,\pm\infty 
,2-\frac{2}{m},2 \left(\frac{1}{m}+1\right) 
m'+4\right\},\\k=+1,~~|\epsilon|\to1^{\pm}\end{array}\right.$&$0$&$-\frac{
\epsilon ^2}{k} $&$-\frac{1}{3}$\\[0.75 ex]
$P^{2}$&$\left(0,0,-6\left[\frac{2 k}{\epsilon ^2-3 
k}+1\right],0\right)$&$\left[\frac{3}{2},-\frac{8 \epsilon ^2}{\epsilon ^2-3 
k}+1,j_1,j_2 \right]$&$\frac{4 \epsilon 
^2}{\epsilon ^2-3 k}$&$0$&$-\frac{8 k}{\epsilon ^2-3 k}-3$\\[0.75 ex]
$P^{3}$&$\left(\frac{2 k}{k-\epsilon 
^2},0,-6,0\right)$&$\left\{\begin{array}{l}\left[-8,-\frac{6 
(m+1)}{m},l_1,l_2 \right],\\ 
(l_1=-3,l_2=3),~~k=-\epsilon ^2\end{array}\right.$&$\frac{8 k}{\epsilon 
^2-k}+4$&$0$&$-3$\\[0.75 ex]
$P^{4}$&$\begin{array}{l}\left(\frac{k (m-1)+(m+1) \epsilon ^2}{m \left[k 
(m+2)+m \epsilon ^2\right]},\frac{(m+1) \left[k (m-1)+(m+1) \epsilon 
^2\right]}{m \left[k (m+2)+m \epsilon ^2\right]}\right.,\\\left.-\frac{6 m}{2 
m+1},0\right)\end{array}$&$\left\{\begin{array}{l}\left[\frac{6}{2 
m+1}-2,3,\frac{3}{2 m+1},\frac{6 (m+1) \left(m'+1\right)}{2 
m+1}\right],\\k=-\epsilon^2\end{array}\right.$&$\frac{2 (m-1) \left(k+\epsilon 
^2\right)}{k (m+2)+m \epsilon ^2}$&$0$&$\frac{2}{2 m+1}-1$\\[0.75 ex]
\hline\hline
\end{tabular}
\caption{The critical points   in the general $f(Q)$ case, namely with    
general $m(r)$. We have defined $j_1=\frac{6 \left[k (m-1)+(m+1) \epsilon 
^2\right]}{m \left(3 k-\epsilon 
^2\right)}$, $j_2=\frac{6 \left[(m+1) \left(k-\epsilon ^2\right) m'+2 k 
m\right]}{m \left(3 k-\epsilon ^2\right)}$, 
$l_1=-\frac{3 \left(-\sqrt{33 k^2-30 k \epsilon ^2+\epsilon ^4}+k+\epsilon 
^2\right)}{4 \left(k-\epsilon ^2\right)}$ and $l_2=\frac{3 \left(\sqrt{33 
k^2-30 
k \epsilon ^2+\epsilon ^4}+k+\epsilon ^2\right)}{4 \left(\epsilon 
^2-k\right)} $.
}
\label{tab1}
\end{table}
\end{center}

The physical properties of these critical points  are the following:

\begin{itemize}\label{items}
\item \textbf { Point $P^{m}$:} It corresponds to dark-matter ($\Omega^m=1$) 
dominated era with total equation-of-state  parameter 
$w^{eff}=0$. Its eigenvalues imply that it is a saddle point and thus it can be 
the intermediate state of the Universe.

\item \textbf {  Point $P^{k}$:} It corresponds to a curvature-dominated era 
and it is a saddle point.

\item \textbf {  Curve of points $P^{ds}$:} It corresponds to a   dark-energy 
dominated Universe (since $\Omega^m=\Omega^k=0$ we have $\Omega^{de}=1$), with  
$w^{eff}=-1$, namely to the de Sitter solution. Although it has a zero 
eigenvalue, application of the center manifold theorem 
\cite{Coley:2003mj,Bahamonde:2017ize} shows that this point is stable and thus 
it can 
attract the Universe at late times.

\item \textbf {  Point $P^{1}$:} This point exists only for non-flat geometry.
It corresponds to a 
curvature-dominated solution if $k=\pm\epsilon ^2$, and it is unstable for 
every values of $m$ and $m'$.

\item \textbf {  Point $P^{2}$:} This point is physical (i.e. having 
$0\leq\Omega^m\leq1$) only for $k=1$ and for $\epsilon^2\leq1$. In 
this case $\Omega^m$ and $\Omega^{de}$ are both between 0 and 1 and thus this 
point can alleviate the coincidence problem. Additionally, it has 
$-1\leq w^{eff}\leq-1/3$ and thus it corresponds to accelerated solution.  The 
fact that it is unstable makes this point a good candidate for the description 
of inflation with a successful exit.

\item \textbf {  Point $P^{3}$:} This point is physical only for $k=-1$ and for 
$1\leq\epsilon^2\leq5/3$, in which case  $\Omega^m$ and $\Omega^{de}$ are both 
between 0 and 1. It corresponds to super-acceleration and it is unstable.

\item \textbf { Curve of points $P^{4}$:} The properties of this curve cannot 
be inferred without specifying $m(r)$, namely the $f(Q)$ form.  

\end{itemize}

In summary, $f(Q)$ cosmology in non-flat Universe exhibits the desired features 
of saddle matter-dominated era and stable late-time dark-energy era. However, 
apart from these, we obtain interesting features that arise solely from 
non-zero curvature, such as a point which can alleviate the coincidence 
problem, or a point that corresponds to a curvature-driven inflation which is 
unstable and thus it can easily acquire a successful inflation exit. 
Nevertheless, since some features cannot be extracted for the general $f(Q)$ 
form, in the following subsection we examine a specific $f(Q)$  case.

\subsection{Application for $f(Q)=\eta Q^{n}$  }\label{sec6}

Let us apply the above general analysis in the case $f(Q)=\eta Q^{n}$. Such a 
choice, according to (\ref{ds1bb}) corresponds to    $m=n-1=const.$ and
  $r=-n=const.$, and thus $x_2=rx_1$, which implies that  variable $x_2$ is not 
needed. We first examine 
the flat case   and then we continue to  $k=\pm1$.

\subsubsection{$k=0$}\label{sec6.1}

\begin{figure}[h!]
\begin{center}
\epsfig{figure=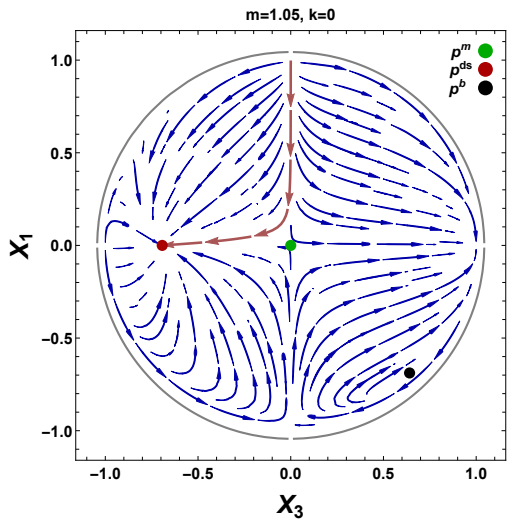,width=7cm}
\caption{{\it{
The phase-space behavior in the specific case of   $f(Q)=\eta Q^{n}$ gravity, 
for $k=0$. The Universe passes through the saddle matter-dominated point  
$p^{m}$ at intermediate times, before it results to the dark-energy dominated de 
Sitter solution $p^{ds}$.  }
}}
\label{fig1}
\end{center}
\end{figure}

In this case   $\Omega^k$ is absent and we acquire    a 
two-dimensional system, namely   (\ref{ds7}) and (\ref{ds9}) with zero 
$\Omega^k$ terms. The corresponding physical critical points are shown in 
Table \ref{tab2}. As we can see, we obtain an unstable dark-matter dominated 
point, namely $p^m$, as well as a stable  dark-energy dominated de Sitter 
solution $p^{de}$. However, for $1\leq m\leq2$ we obtain point $p^{b}$, in 
which   $\Omega^m$ and $\Omega^{de}$ are both between 0 and 1 and thus this 
point can alleviate the coincidence problem, while it  has 
$-3/5\leq w^{eff}\leq-1/3$. Note that this point for $1\leq m$  is unstable.
In Fig. \ref{fig1} we depict the critical 
points, using for convenience the  new variables 
$x\equiv X/\sqrt{1-X^2-Y^2}$ and $y\equiv Y/\sqrt{1-X^2-Y^2}$  in order to 
compactify them.

\begin{center}
\begin{table}[h]
\centering
\begin{tabular}{l @{\hskip 0.1in} l@{\hskip 0.1in} l @{\hskip 0.1in}l @{\hskip 
0.1in}l @{\hskip 0.1in}} \hline\hline
Fixed point     &Coordinates $(x_1,x_3)$           &Eigenvalues  &$\Omega^m$    
&$w^{eff}$\\[0.5 ex]
\hline
$p^{m}$&$\left(0,0\right)$&$\left[3,-\frac{3}{2}\right]$&$1$&$0$\\[0.75 ex]
$p^{ds}$&$\left(-\frac{1}{m},0\right)$&$\left[-3,-3\right]$&$0$&$-1$\\[0.75 ex]
$p^{b}$&$\left(\frac{m+1}{m^2},-\frac{6 m}{2 m+1}\right)$&$\left[-\frac{3 
\left(-2 m^2+\sqrt{1-4 m^2 (m+1) (7 m+5)}+1\right)}{4 m (2 m+1)},\frac{3 
\left(2 m^2+\sqrt{1-4 m^2 (m+1) (7 m+5)}-1\right)}{4 m (2 
m+1)}\right]$&$2-\frac{2}{m}$&$\frac{2}{2 m+1}-1$\\[0.75 ex]
\hline\hline
\end{tabular}
\caption{The critical points   for the case $f(Q)=\eta Q^{n}$, with 
$k=0$.}
\label{tab2}
\end{table}
\end{center}

\subsubsection{$k=-1$}\label{sec6.2}

In the case of $k\neq0$ the system of 
dynamical equations contains  (\ref{ds7}), (\ref{ds9}) and (\ref{ds10}). The 
critical points are presented in     Table~\ref{tab3}. As can be 
seen there is the unstable  dark-matter dominated point  $Q^{m}$, and the 
unstable    curvature-dominated 
point $Q^{k}$. Additionally, there exist a stable dark-energy dominated de 
Sitter solution  $Q^{ds}$. Moreover, similarly to  points $P^{4}$ in 
Table~\ref{tab1} and $p^b$ in
Table~\ref{tab2} above, there is a point 
$Q^{3}$
in 
which   $\Omega^m$ and $\Omega^{de}$ are both between 0 and 1 and thus this 
point can alleviate the coincidence problem.

\begin{center}
\begin{table}[h]
\centering
\begin{tabular}{l @{\hskip 0.1in} l@{\hskip 0.1in} l @{\hskip 0.1in}l @{\hskip 
0.1in}l @{\hskip 0.1in}l}\hline\hline
Fixed point     &Coordinates $(x_1,x_3,\Omega^{k})$           &Eigenvalues  
&$\Omega^m$   &$\Omega^k$   &$w^{eff}$\\[0.5 ex]
\hline
$Q^{m}$&$\left(0,0,0)\right)$&$\left[3, -\frac32, 1\right]$&$1$&$0$&$0$\\[0.75 
ex]
$Q^{k}$&$\left(0,0,1\right)$&$\left[-2,-1,2\right]$&$0$&$1$&$-\frac13$\\[0.75 
ex]
$Q^{ds}$&$\left(-\frac{1}{m},0,0\right)$&$[-3,-3,-2]$&$0$&$0$&$-1$\\[0.75 ex]
$Q^{1}$&$\left(0,-6\left[\frac{2 k}{\epsilon ^2-3 
k}+1\right],0\right)$&$\left[\frac{3}{2},-\frac{8 \epsilon ^2}{\epsilon ^2-3 
k},\frac{6 \left[k (m-1)+(m+1) \epsilon ^2\right]}{m \left(3 k-\epsilon 
^2\right)}\right]$&$\frac{4 \epsilon ^2}{\epsilon ^2-3 k}$&$0 $&$-\frac{8 
k}{\epsilon ^2-3 k}-3$\\[0.75 ex]
$Q^{2}$&$\left(0,-2,-\frac{\epsilon 
^2}{k}\right)$&$\left[2-\frac{2}{m},1,2\right]$&$0$&$-\frac{\epsilon 
^2}{k}$&$-\frac{1}{3}$\\[0.75 ex]
$Q^{3}$&$\left(\frac{k (m-1)+(m+1) \epsilon ^2}{m \left[k (m+2)+m \epsilon 
^2\right]},-\frac{6 m}{2 
m+1},0\right)$&$\left\{\begin{array}{l}\left[\frac{6}{2 m+1}-2,3,\frac{3}{2 
m+1}\right],\\k=-\epsilon^2\end{array}\right.$&$\frac{2 (m-1) \left(k+\epsilon 
^2\right)}{k (m+2)+m \epsilon ^2}$&$0$&$\frac{2}{2 m+1}-1$\\[0.75 ex]
\hline\hline
\end{tabular}
\caption{The critical points   for the case $f(Q)=\eta Q^{n}$, with 
$k=\pm1$.}
\label{tab3}
\end{table}
\end{center}

In the presence of   spatial curvature the two points $Q^{1}$ and $Q^{2}$ 
appear, too. $Q^{1}$ corresponds to $P^{2}$ of 
Table~\ref{tab1}, and it is  is physical   only for $k=1$ and for 
$\epsilon^2\leq1$. It has both $\Omega^m$ and $\Omega^{de}$     between 0 
and 1 and thus this point can alleviate the coincidence problem. Additionally, 
it has $-1\leq w^{eff}\leq-1/3$ and therefore it corresponds to accelerated 
solution.  The fact that it is unstable makes this point a good candidate for 
the description of inflation. Furthermore, point   $Q^{2}$ corresponds to  
$P^{1}$, namely it describes   a curvature-dominated solution if $k=\pm\epsilon 
^2$, and it is unstable.
\begin{figure}[h!]
\begin{center}
\epsfig{figure=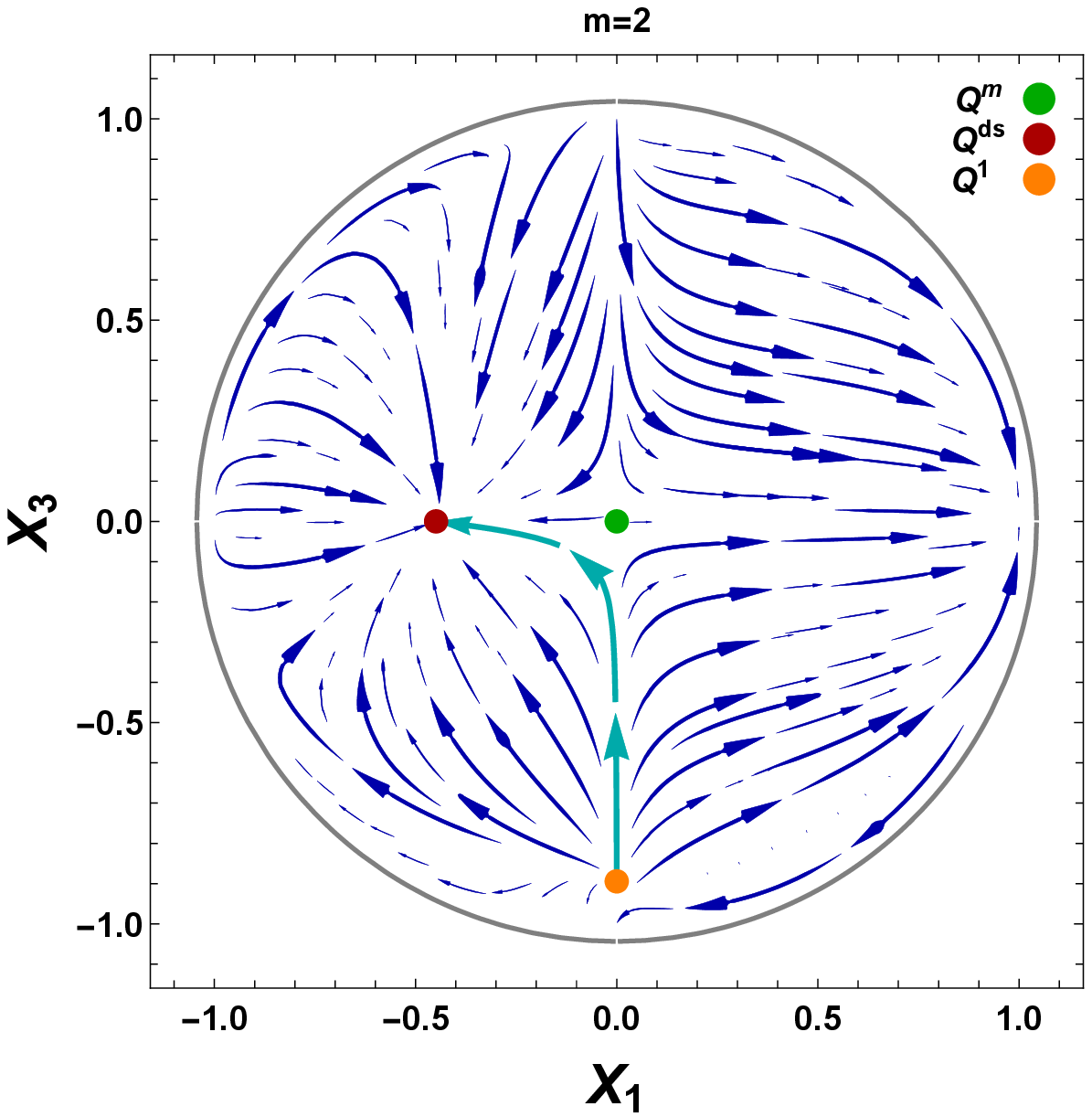,width=7cm}\hspace{2mm}
\epsfig{figure=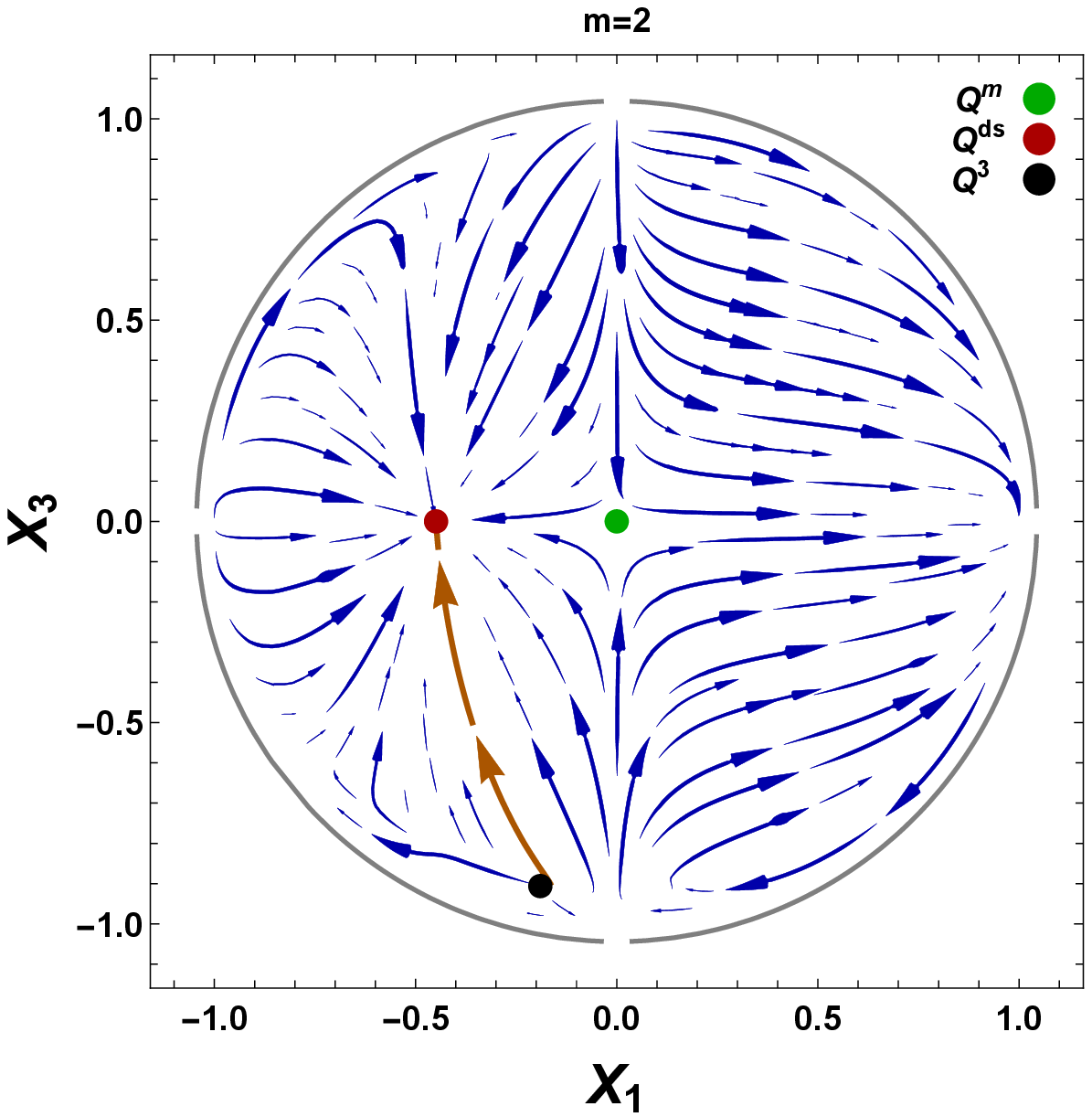,width=7cm}
\caption{{
\it{The phase-space behavior in the specific case of   $f(Q)=\eta Q^{n}$ 
gravity, 
for $k=-1$. 
Left panel: for the choice  $\epsilon=0.1$ the system 
starts from the inflationary point $Q^{1}$, then it passes close to the 
matter-dominated point  $Q^{m}$ and finally it results to the dark-energy 
dominated de Sitter solution  $Q^{ds}$. 
 Right panel: for the choice  $\epsilon=1$  the system starts from the scaling 
point $Q^{3}$ and it results to the dark-energy 
dominated de Sitter solution  $Q^{ds}$ without passing sufficiently close to the
  matter-dominated point  $Q^{m}$.  }
}}\label{fig2}
\end{center}
\end{figure}

In order to present the above features in a more transparent way, we proceed to 
numerical investigation and in  Fig.~\ref{fig2} we depict the corresponding 
phase-space behavior in the $X_1-X_3$ plane. In the left  panel the system 
starts from the inflationary point $Q^{1}$, then it passes close to the 
matter-dominated point  $Q^{m}$ and finally it results to the dark-energy 
dominated de Sitter solution  $Q^{ds}$. As we can see, the role of spatial 
curvature is crucial in obtaining such a thermal 
history of the Universe.
Moreover, for completeness, in the  right panel of 
Fig.~\ref{fig2} we present a parameter-case in which    the trajectories 
starting from $Q^{3}$ do not   
 approach   $Q^{m}$ efficiently.
 
 \begin{figure}[h!]
\begin{center}
\epsfig{figure=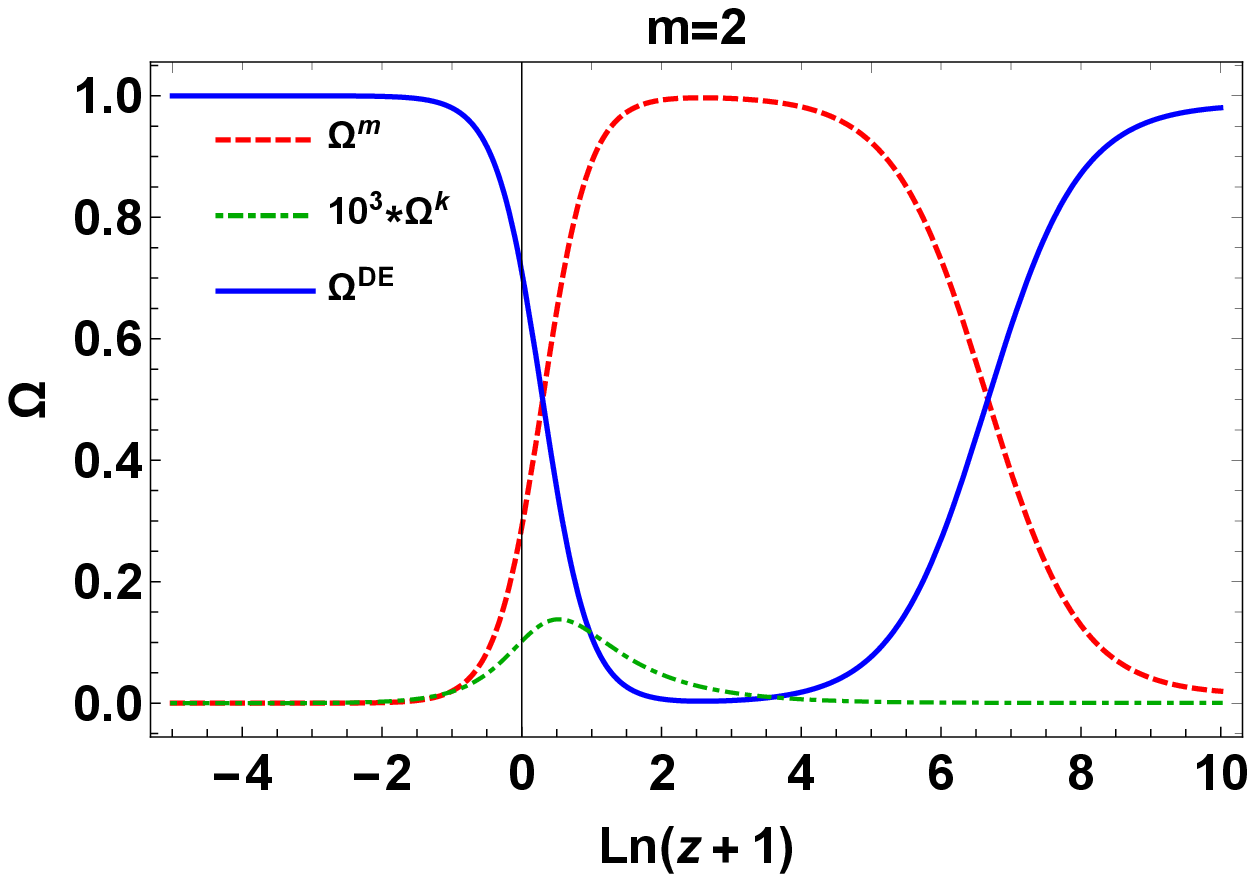,width=7.cm}\hspace{2mm}
\epsfig{figure=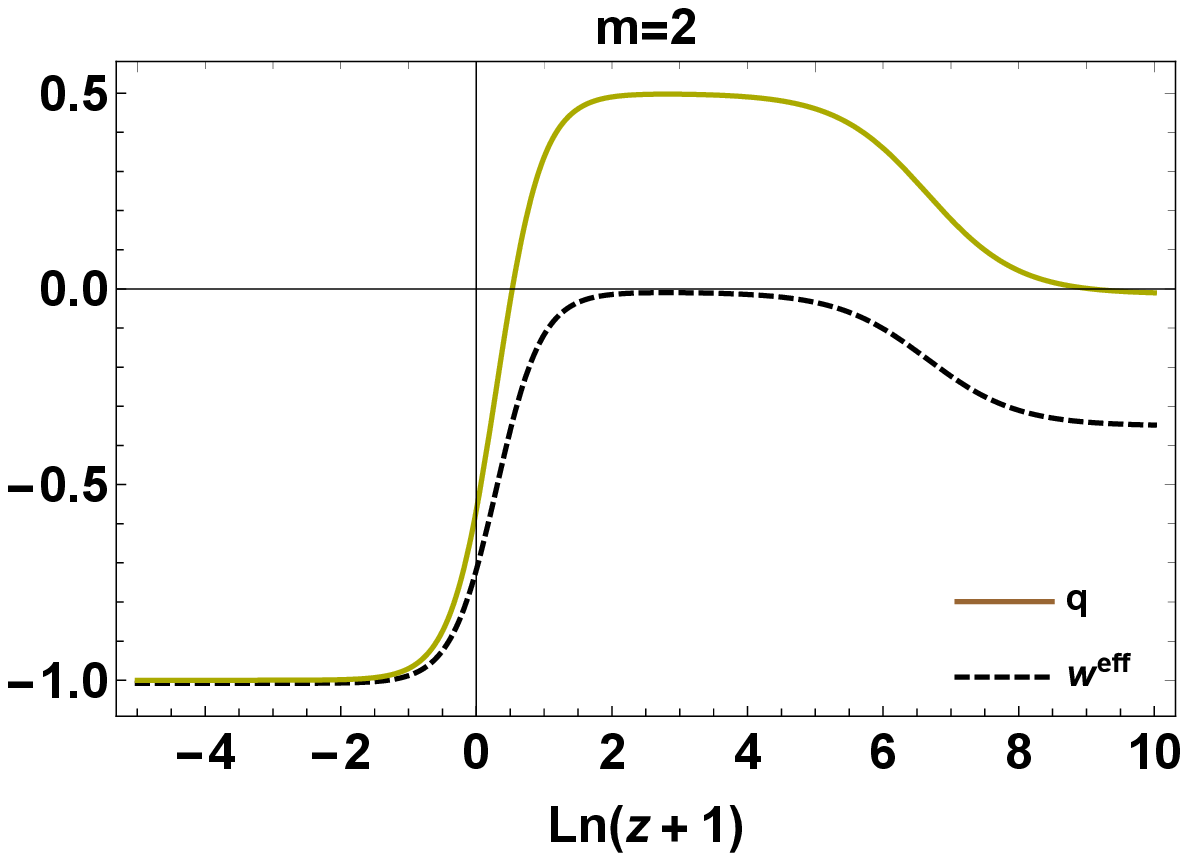,width=7.cm}
\caption{
{\it{The redshift evolution of the density parameters (left panel) and of the 
deceleration and equation-of-state parameters (right panel)
in the 
specific case of   $f(Q)=\eta Q^{n}$ 
gravity, 
for $k=-1$ and $\epsilon=0.1$
We have set  the initial conditions 
$x_{1i}=-1.05\times10^{-10}$, $x_{3i}=-2.00013$ and 
$\Omega^{k}_{i}=5\times10^{-7}$.}}}
\label{fig3}
\end{center}
\end{figure}

 In the left panel of Fig.~\ref{fig3} we   provide the 
redshift-evolution of the density parameters, while in the right panel we 
depict the evolution of the deceleration and total equation-of-state  
parameters (note that $\ln(1+z)=-\ln a=-N$).  Interestingly enough, we  observe 
a transition from the initial accelerated expansion stage, to the intermediate 
matter-dominated non-accelerating era, and then to the 
 final accelerated expansion phase. Additionally, note that  
  the spatial 
curvature density parameter grows when the domination of the matter and the 
dark energy phases is reversed.

\begin{figure}[h!]
\begin{center}
\epsfig{figure=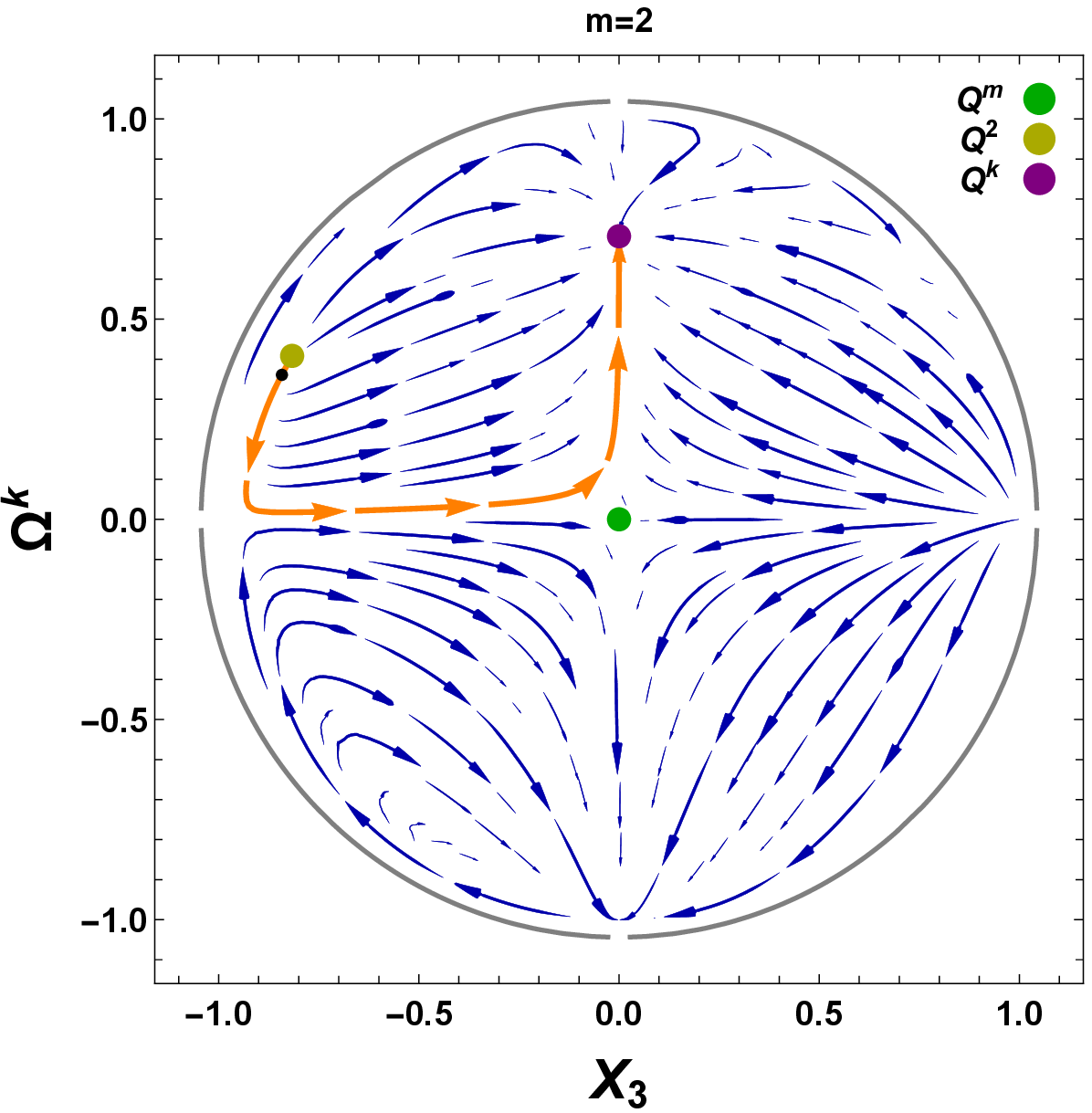,width=7.cm}\vspace{2mm}\\
\epsfig{figure=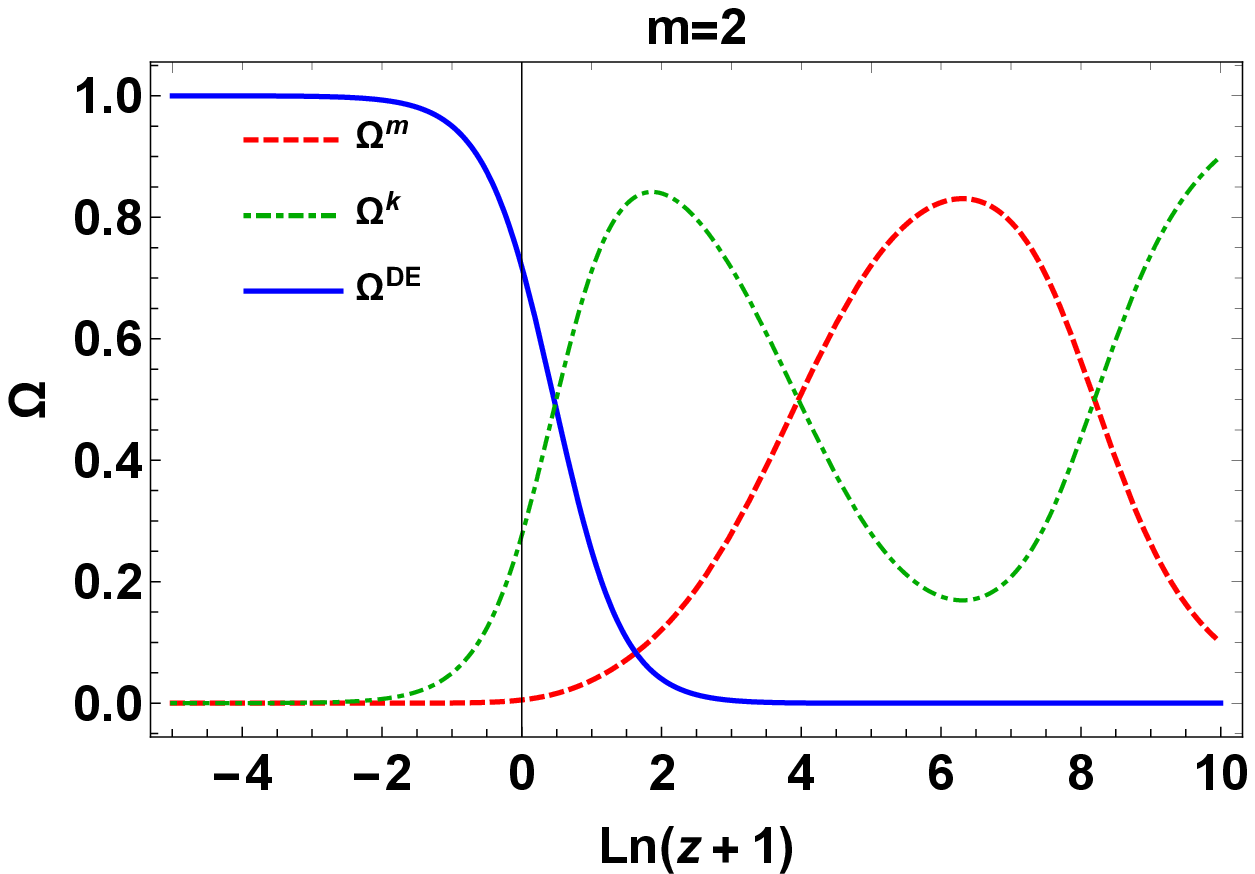,width=7cm}\hspace{2mm}
\epsfig{figure=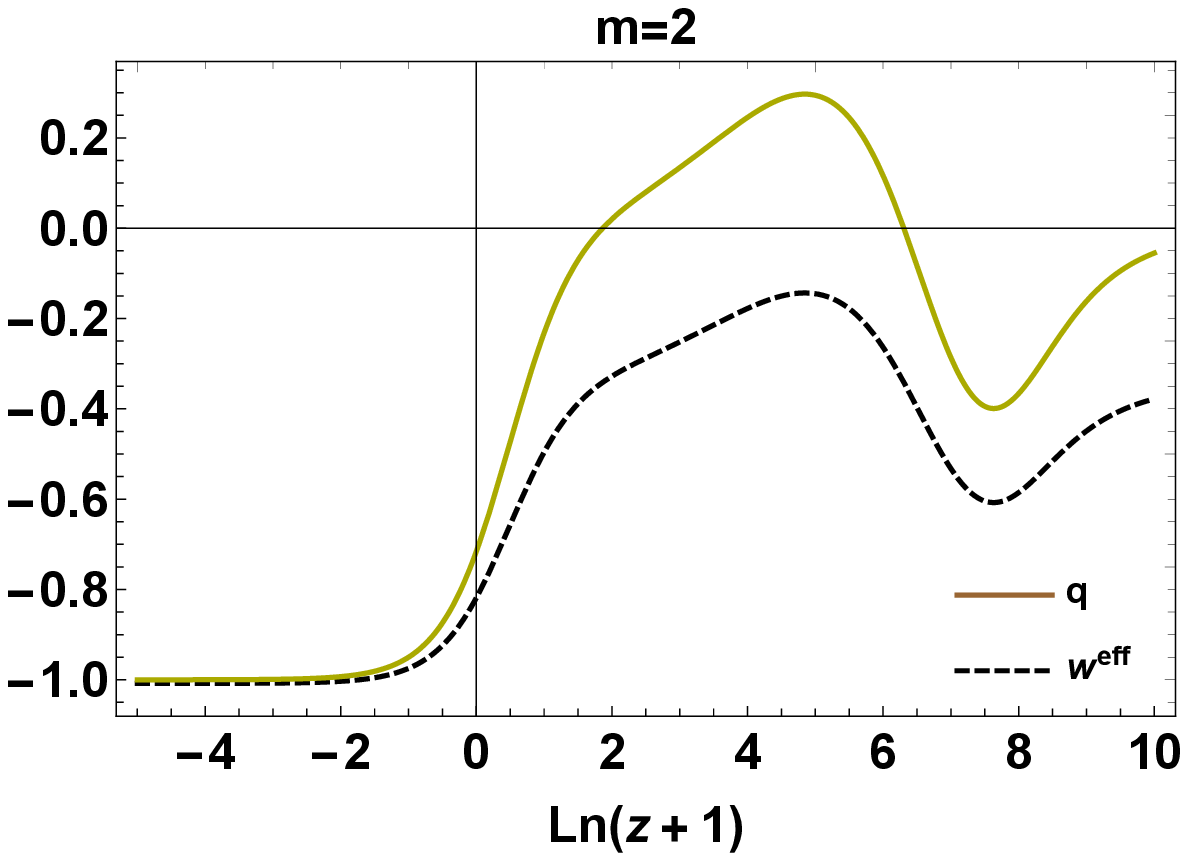,width=7cm}
\caption{
{
\it{
Upper panel:
the phase-space behavior in the specific case of   $f(Q)=\eta 
Q^{n}$ gravity, in the $x_1$-plane,
for $k=-1$  and $\epsilon=1$. The orange curve represents the transition from 
the curvature-dominated point $Q^{2}$ to the 
matter-dominated point  $Q^{m}$ and then to the 
curvature-dominated solution  $Q^{k}$. 
  Lower panels: 
the corresponding redshift evolution of the density parameters  and 
of the 
deceleration and equation-of-state parameters.
We have set  the initial conditions 
$x_{1i}=-1.5\times10^{-7}$, $x_{3i}=-2.1$ and 
$\Omega^{k}_{i}=0.9$, related to the black dot near $Q^{2}$ in the upper 
panel.
  }
}
}\label{fig4}
\end{center}
\end{figure}

Finally, in order to illustrate the behavior of the phase-space trajectories 
near the curvature-dominated points $Q^{k}$ and $Q^{2}$, we focus on  
the $x_1=0$ plane.    In  upper panel  of Fig.~\ref{fig4} we display
$Q^{m}$, $Q^{k}$ and $Q^{2}$ in the $x_1=0$ plane for $k=-1$. Note that in the 
$x_1=0$ plane one cannot indicate the point $Q^{ds}$ for which one acquires 
$x_1=-1/m$. As we can see, for particular initial values the dark-matter 
dominated phase falls between two different epochs with considerable values of 
$\Omega^{k}$. In particular, the Universe evolves from an epoch with 
curvature domination   to the dark-matter dominated era and then to another 
curvature-dominated epoch. The lower panels  of Fig.~\ref{fig4} show the time 
evolution of the density parameters, and  the deceleration and total 
equation-of-state  parameters. As we can see,  the transition
$Q^{2}$-$Q^{m}$-$Q^{k}$-$Q^{ds}$ is also possible for a particular set of the 
initial values.

\subsubsection{$k=+1$}\label{sec6.3}

\begin{figure}[h!]
\begin{center}
\epsfig{figure=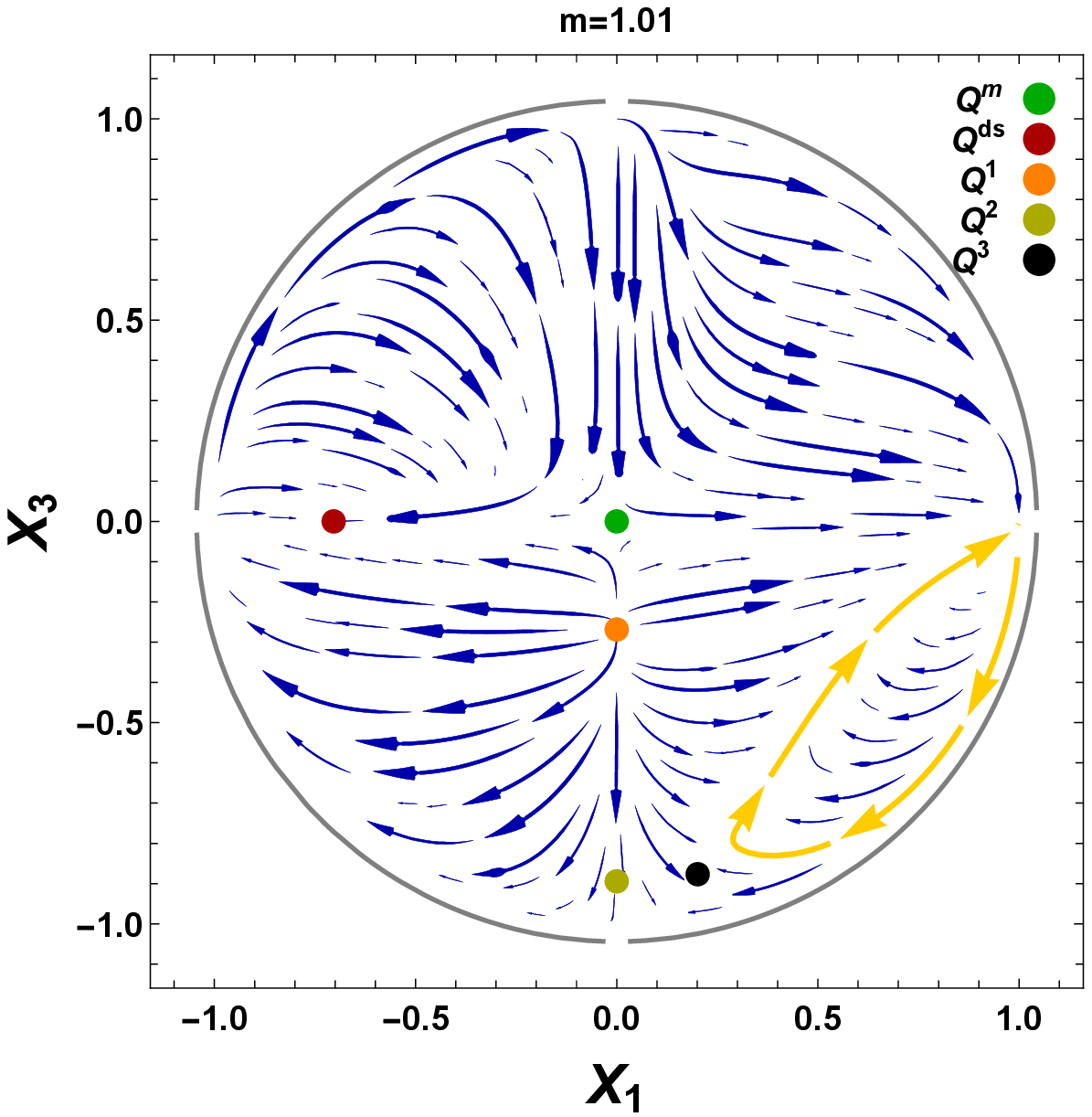,width=7cm}\hspace{2mm}
\epsfig{figure=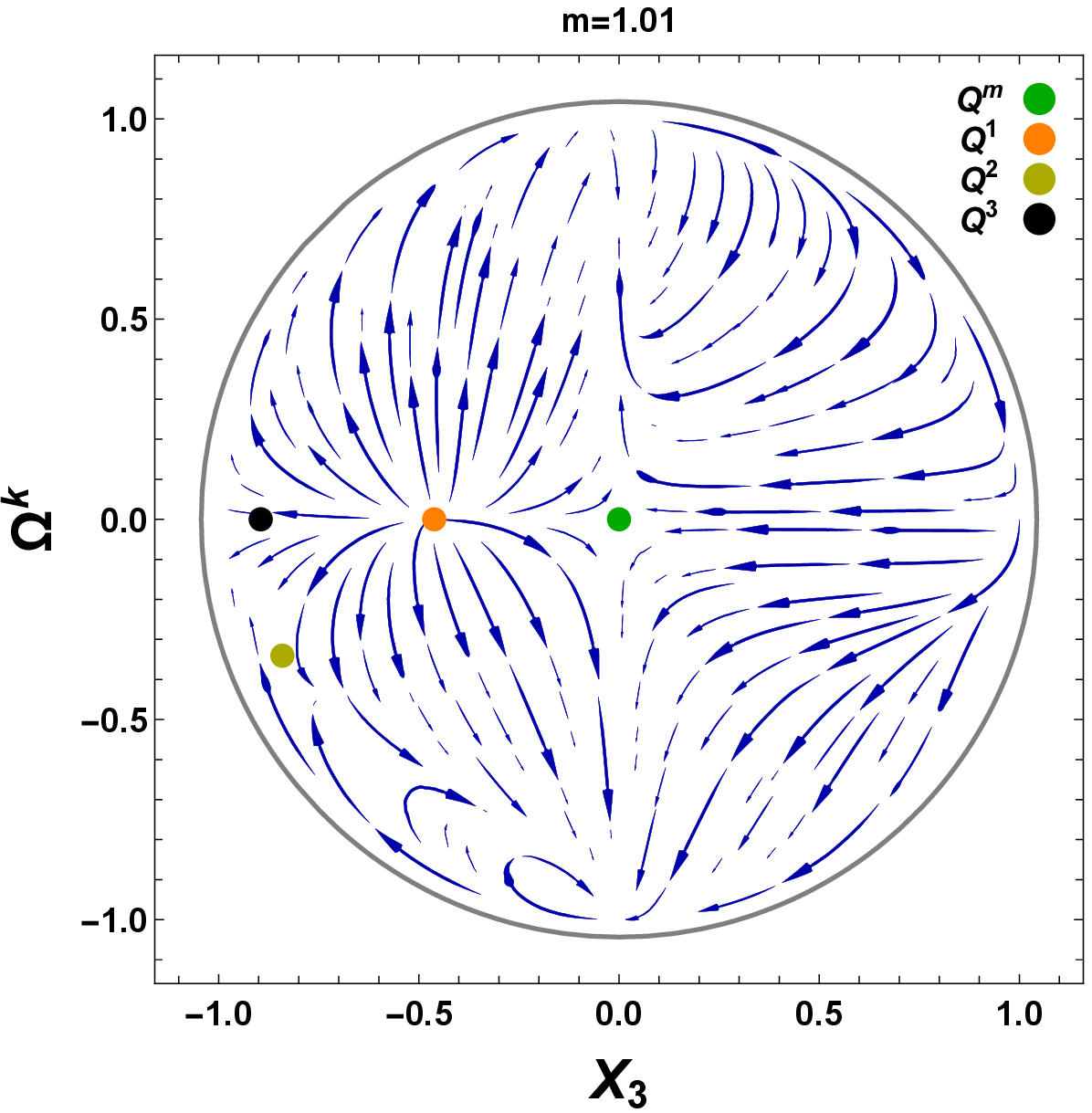,width=7cm}
\caption{
{
\it{The phase-space behavior in the specific case of   $f(Q)=\eta Q^{n}$ 
gravity, 
for $k=+1$, in the $X_1-X_3$ plane (left panel) and in the $Q^{k}-X_3$ plane 
(right panel).  The unstable  $Q^{1}$ point stands in between 
$Q^{m}$ and $Q^{2}$/$Q^{3}$ and thus it blocks transitions between them,
and hence we can only obtain the 
usual transition from   matter to dark-energy dominated phases.
 }}}\label{fig6}
\end{center}
\end{figure}

In the case of $k=+1$ the system of 
dynamical equations    (\ref{ds7}), (\ref{ds9}) and (\ref{ds10}) exhibits the 
critical points   presented in     Table~\ref{tab3}.  In particular, one 
has points $Q^{m}$ and $Q^{ds}$, however in this case  $Q^{k}$ and $Q^{1}$ are 
not physical.    Point $Q^{2}$  corresponds to a closed spatial curvature 
dominated era. Finally, 
$Q^{3}$ exists, in 
which   $\Omega^m$ and $\Omega^{de}$ are  between 0 and 1 and thus  it  can 
alleviate the coincidence problem.  In Fig.~\ref{fig6} we   plot the phase-space 
trajectories  in both $\Omega^{k}=0$ and $x_1=0$ planes. As we can see, we 
cannot obtain any transition between $Q^{m}$ and $Q^{2}$/$Q^{3}$. Hence, we 
conclude that under positive spatial curvature we can only obtain the usual 
transition from   matter to dark-energy dominated phases.


\section{Concluding remarks}
\label{sec8}

In this manuscript we investigated the cosmological implications of $f(Q)$ 
gravity, which is a modified theory of gravity based on non-metricity, in 
non-flat FLRW geometry. After presenting the relevant cosmological equations, 
we performed a detailed dynamical-system analysis in order to reveal the global 
features of the evolution, independently of the initial conditions. 

Firstly, we performed the analysis keeping the $f(Q)$ function  completely 
arbitrary. As we showed, the cosmological scenario admits a dark-matter 
dominated point, which is saddle and  thus it can be 
the intermediate state of the Universe, as well as dark-energy 
dominated  de Sitter solution which is stable and thus 
it can attract the Universe at late times. 
However, the main result of the 
present work is that there are additional critical  points and curves of 
critical points which exist solely due to curvature

In particular, we found that there are points which are curvature-dominated and 
correspond to accelerating expansion, and the fact that they are unstable 
makes them  good candidates for the description of inflation. Additionally, 
there is a point in which the dark-matter and dark-energy density parameters 
are both between zero and one, and thus  it  can alleviate the coincidence 
problem. Finally, there is a saddle point which is completely dominated by 
curvature.

In order to provide a specific example, we applied our general analysis to the 
power-law case  $f(Q)=\eta Q^{n}$. In this specific model, the Universe 
exhibits the general features presented above, namely a saddle matter-dominated 
point and a late-time dark-energy dominated attractor. Furthermore, it has 
 points that exist only in the non-flat case, which can alleviate the 
coincidence problem, as well as curvature-dominated accelerating unstable points 
that can describe the early-time inflationary epoch. In this case we performed 
a numerical investigation showing that the system    in the open 
geometry case exhibits a transition from 
the initial accelerated expansion stage, to the intermediate matter-dominated 
non-accelerating era, and then to the 
 final accelerated expansion phase, while the curvature density parameter 
exhibits a peak at intermediate times.

In summary, $f(Q)$ cosmology in non-flat Universe exhibits the desired behavior 
known from the flat case, however it additionally exhibits qualitatively novel 
features that arise solely from non-zero curvature. This fact, alongside 
possible indications that non-zero curvature could alleviate the cosmological 
tensions, makes it both interesting and necessary to further investigate 
modified gravity, and in particular $f(Q)$ gravity, in non-flat geometry.

\begin{acknowledgments}
 This research was partially supported by the UTAR Research Fund Scheme. ENS would like to acknowledge the contribution of the COST Action 
CA21136 ``Addressing observational tensions in cosmology with systematics and 
fundamental physics (CosmoVerse)''. 
\end{acknowledgments}



\end{document}